\documentclass{revtex4}% no class options needed by now
\usepackage[english]{babel}

\usepackage{graphicx}
\usepackage{amssymb,amsfonts,amsmath,esint}
\usepackage[latin1]{inputenc}
\usepackage{epstopdf}
\def \r{{\bf r}}
\def \R{\mathbb{R}}
\def \N{\mathbb{N}}
\def \Z{\mathbb{Z}}
\def \C{\mathbb{C}}
\def \bq{\begin{equation}}
\def \eq{\end{equation}}
\newcommand{\ket}[1]{\left | {#1} \right.\rangle}

\begin{document}

\title{All-optical photonic band control in a quantum metamaterial}
% \subtitle{subtitle}
\author{D. Felbacq\footnote{Corresponding author\quad E-mail:~\textsf{didier.felbacq@umontpellier.fr}}, E. Rousseau}
\affiliation{University of Montpellier, Laboratory Charles Coulomb UMR CNRS-UM 5221, Place Bataillon, B\^at. 21 CC074, 34095 Montpellier Cedex 05, France}
%\address[2]{Affiliation and address of Second X. Author and Third Y. Author}
%\shortauthors{D. Felbacq et al.}
\begin{abstract}
Metamaterials made of periodic collections of dielectric nanorods are considered theoretically. When quantum resonators are embedded within the nanorods, one obtains a quantum metamaterial, whose electromagnetic properties depend upon the state of the quantum resonators. The theoretical model predicts that when the resonators are pumped and reach the inversion regime, the quantum metamaterial exhibits an all-optical switchable conduction band. The phenomenon can be described by considering the pole stucture of the scattering matrix of the metamaterial.
\end{abstract}
\maketitle
% \noindent

\section{Introduction}
Metamaterials are mesoscopic structures made on a set of basic cells, each basic cell containing resonant elements \cite{capolino}. They are considered in the regime when they are illuminated by an incident field whose wavelength in vacuum is much larger than the period. In that case, they behave as homogeneous materials, whose properties are not found in natural materials. While they have been intensely studied as 3D structures \cite{review}, they are also studied in the guise of 2D materials, where resonant elements are deposited on a surface. These structures are called "metasurfaces" \cite{metasurf}. There is a huge literature devoted to passive metamaterials and metasurfaces, covering as various domains as super-lensing, invisibility cloak, directional propagation, phase control...Metamaterials are generally made on metallic elements, in which case they are lossy, especially in the higher frequency regions of the spectrum. It was then imagined to insert gain inside metamaterials, under the form of quantum dots for instance \cite{gain}, in order to compensate for the losses inherent to the metals. Quantum dots have also been inserted inside metamaterials to make them active and able to emit light \cite{metaQD}. Another different trend concerns quantum metamaterials \cite{rewalex,moiQ}. The term was first coined by A. M. Zagoskin in the context of superconducting metamaterials \cite{Qmat}, where structures are made on arrays of Josephson junctions. They are quantum structure in the sense that the quantum state of the basic cells (e.g. Josephson junctions) influences the collective behavior of the entire structure. It was for instance demonstrated that, upon preparing the junctions in a superposition of states, it was possible to open or close a photonic band gap \cite{Qmat2,manipQmat}.

In the present work, we present a metamaterial made of a periodic collection of dielectric resonators in which a quantum oscillator (denoted QO in the following) is inserted. 
The geometry at stake here is much more complicated than the textbook 1D cavity usually dealt with theoretically in quantum optics. We do provide a treatment essentially based on the scattering matrix non-perturbative approach \cite{cohen}, in order to investigate the various effects that could be expected to exist in such structures.
First, the phenomenology for one scatterer with a QO inserted is presented, then the collective behavior of a finite periodic set of such scatterers is investigated and it is shown that it is possible to open and close a conduction band according to the state of the oscillators.

\section{A scatterer with a quantum emitter embedded}
In order to understand the behavior of the scatterers with a QO, let us consider first the case of a single scatterer. 
The spectrum of the nanorods comprises a discrete set of inner pure modes $u_n(\r,\omega_n)$ with eigenfrequencies $\omega_n$ and a continuum of scattering modes $u(\r;\omega)$.
Formally, a general solution of Maxwell equations in the presence of the structure can therefore be written, using the generalized spectral theorem \cite[p.221]{reedsimon}:
%\bq
%u(\r)=\int dk a(k) u(\r;k) +\sum_{np} \int dk a(\omega) u_{np}(\r;\omega).
%\eq
%We are mainly interested in the low frequency modes, and that these modes are very weakly dispersive. From this we deduce that one of the quantum number can be skipped (i.e. we keep only electric Mie resonances) and the integral is supported by $\delta(k)$, this gives:
\bq
u(\r)=\int d\omega b(\omega) u(\r;\omega) +\sum_{n} a_{n} u_{n}(\r,\omega_n).
\eq
By applying the canonical quantization scheme, we obtain the electromagnetic quantum field in the form:
\begin{eqnarray*}
\hat{u}(\r)=\int_0^{+\infty} d\omega \left[ \hat{b}(\omega) +\hat{b}^+(\omega)\right] u(\r;\omega)+
\sum_{n} \left[ \hat{a}_{n} + \hat{a}^+_{n}\right]  u_n(\r,\omega_n)
\end{eqnarray*}
where the annihilation operators $(\hat{b}$ , $\hat{a}_n)$ and the creation operators $(\hat{b}^+$, $\hat{a}^+_n)$ act on the bosonic Fock space ${\cal F}_{\rm bosonic}$ and satisfy the commutation relations: $[\hat{b}(\omega),\hat{b}^+(\omega')]=\delta(\omega-\omega')$ and $[\hat{a}_n,\hat{a}^+_p]=\delta_{np}$.
The pure modes can be described as harmonic oscillators, hence associated with pure point spectrum embedded in the continuous spectrum \cite{reedsimon}, but coupled to the continuum of scattering states. This can be described by an input/output-like formalism \cite{milburn,carmichael} by introducing a coupling potential of the form $V=\sum_n \int d\!\omega \, g_n(\omega)\left[ \hat{a}_n \hat{b}^+(\omega)+\hat{a}_n^+\hat{b}(\omega)\right]$. This coupling induces a radiative shift of the energy of the oscillator and hence a finite lifetime. This results into the existence of quasi-modes associated to complex frequencies $\Omega_p=\omega_p+i\Gamma_p$.
Therefore, as we are
interested in the field that goes outside the structure, it is relevant to use this intput/output-like formalism  in order to obtain a scattering theory between the operators $(H_0,H_0+V)$, where $H_0$ denotes the Hamiltonian corresponding to the continuum of scattering states and $H_0+V$ the hamiltonian corresponding to the continuum of scattering modes plus the pure modes. Let $S(\omega)$ denotes the scattering matrix for the continuum of scattering states. This can be computed by means of the Feshbach projection method \cite{cohen} and leads to the fact that the complex energies of the quasi-modes are poles of $S(\omega)$ \cite{rotter,collision}.
It is important to note that because of the existence of two different conventions for the time dependence, i.e. $e^{\pm i \omega t}$, due in fact to the use of a square root, the scattering matrix is defined on a two-sheeted Riemann surface. The two determinations of the scattering matrix are denoted $S^{\pm}$. Note that the reality of the fields imposes: $S^{\pm}(-\omega)=\overline{S^{\mp}(\overline{\omega})}$. 

The behavior of a scatterer can be understood in terms of poles of the scattering matrix, the so-called scattering resonances \cite{melrose,moipole,moipole2,simon,moiseyev}. For a purely passive scatterer (i.e. dielectric without gain) the electromagnetic behavior is therefore linked to the existence of the so-called Mie resonances. As explained above, these can be understood as open cavity modes of the nanorods, characterized by a set of complex frequencies $\Omega_n=\omega_n+i \Gamma_n$, where $\Gamma_n$ represents the lifetime of the cavity mode. This will exemplified in section \ref{numerical}), where the pole structure of the 
scattering matrix will be computed. This modes can be represented using the formalism of Gamow vectors \cite{bohm}
or quasi-normal modes \cite{hugonin}. 

\begin{figure}
  \includegraphics[width=10cm]{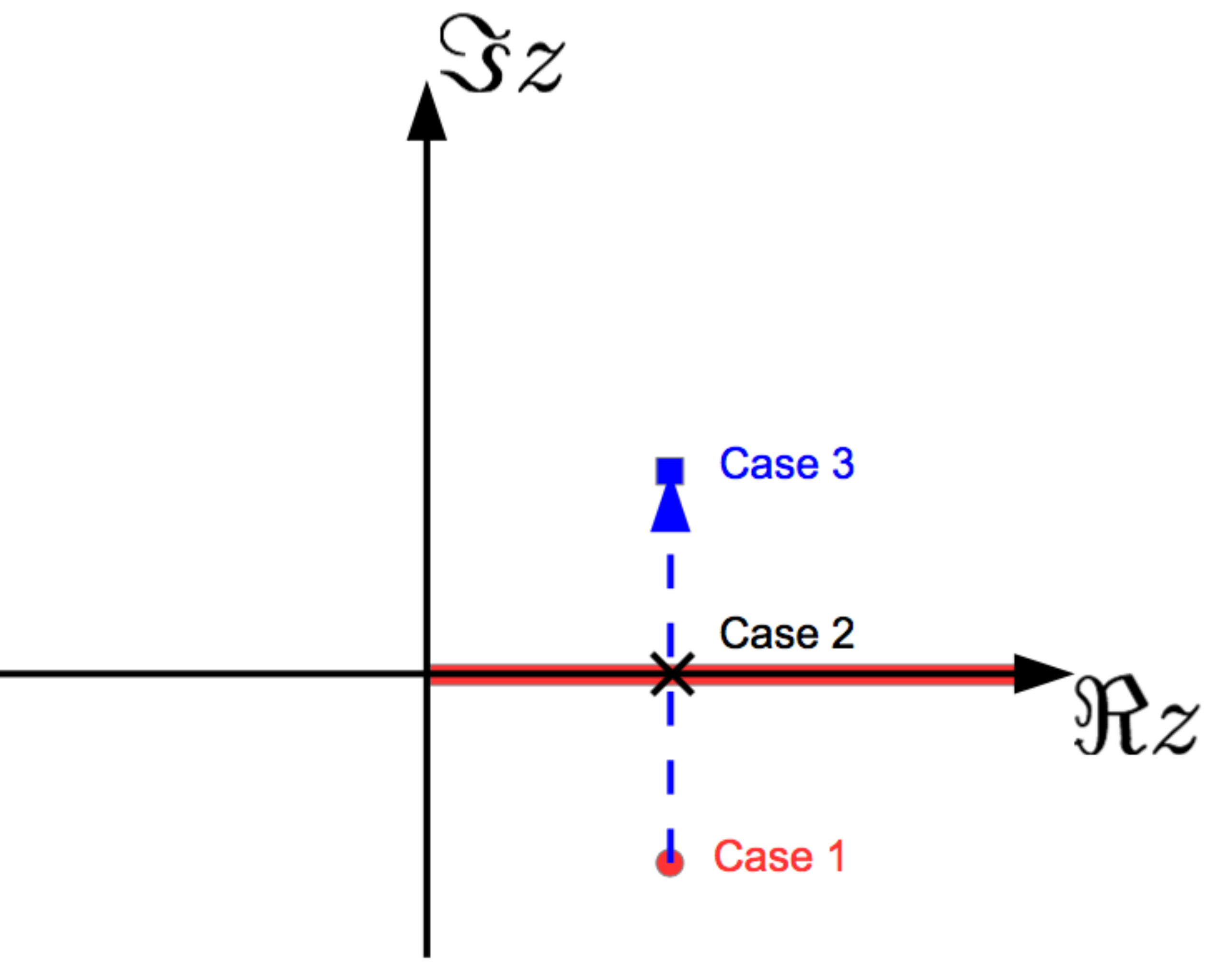}
  \caption{\label{ComplexPlane}Illustration of the pole behavior in the complex plane when some gain is added to the system. Three cases happen. In the first case, the pole lies in the lower part of the complex plane leading to a finite lifetime mode. For the second case, the pole lies in the real axis which is also a branch cut (red bold line). In the third case, the pole lies in the upper part of the complex plane. The corresponding quasi-mode is exponentially amplified.}
\end{figure}

Choosing the $e^{-i\omega t}$ convention implies that the physical sheet is the one where the unstable resonance (i.e. those leading to an exponential decreasing in time for a non-amplifying medium) are situated in the lower part of the complex plane. Choosing this convention, we now suppose that some gain is introduced uniformly inside the scatterer without specifying, for the moment, the physical mechanism at stake.
The introduction of gain induces a shift of the poles towards the upper part of the complex plane of frequencies. 
Three situations can then happen, as to the modes. First, if the gain is small enough, then the considered mode has still a finite, if longer, lifetime. This means that the corresponding mode of the scattering matrix is shifted towards the real axis. 
The second situation corresponds to a sufficient gain that the pole is shifted exactly on the real axis, leading to an unstable point. In this situation, the radiation losses of the quasi-mode are exactly compensated. This means that an eigenvalue (belonging to the pure point spectrum) is embedded in the continuous spectrum. This is generally considered to correspond to the lasing instability \cite{soukoulis}. The fact that the scattering problem is ill-posed at this point is physically irrelevant, as it will be shown in the following by considering the temporal behavior of the field (cf. eq. (\ref{temporal})). The third situation corresponds to a gain high enough that a quasi-mode is now amplified. The pole of the scattering matrix is now in the upper part of the complex plane. The behavior of the pole that has been just described is sketch up in fig.\ref{ComplexPlane}. Note that when the pole goes from the lower part of the complex plane to the upper part it crosses a branch cut. This crossing has important consequences on the time behavior of the field since the pole has to belong to the correct sheet.

The temporal behavior of the field can be obtained by considering the Fourier-Laplace transform of the field in the form: $U(\r,t)=\int S^{\pm}(\omega) A^{\pm}(\r,\omega) e^{\pm i\omega t} d \omega$, where $S(\omega)$ is the scattering matrix and $A(\r,\omega)$ the spectrum of the incident field. 
The reality of the field implies that one can write ($\Re$ denotes the real part):
\bq\label{temporal}
U(\r,t)=2\Re \int_0^{+\infty} A^{\pm}(\r,\omega)S^{\pm}(\omega) e^{\pm i\omega t} d \omega.
\eq
Expanding the meromorphic scattering matrix in Laurent series, and considering that the support of the incident field is in the vicinity of the resonance, we obtain an expression of the form: $$U(\r,t)\sim\int \frac{r^{\pm}(\r,\omega)}{\omega-\omega_0} e^{\pm i\omega t} d \omega.$$ 
Let us now change our point of view and consider this expression as a function of the complex variable $\omega$ (we fix $t$ and $\r$ and make them implicit). We define 
\begin{eqnarray}\label{calU}
{\cal U}^{+}(z)=\int_{\R^+}  \frac{r^{+}(\r,\omega)}{\omega-z}  e^{ i \omega t} d \omega, \Im z >0 \label{calU1}\\
{\cal U}^{-}(z)=\int_{\R^+}  \frac{r^{-}(\r,\omega)}{\omega-z} e^{ -i \omega t} d \omega, \Im z <0.\label{calU2}
\end{eqnarray}
The real field is of course given by: $U=2\Re\left( {\cal U}^{\pm}\right)$. 

In passive media, the poles are situated in the half-plane: $\C^-=\left\{z,\, \Im z <0\right\}$ for the determination ${\cal U}^{-}$ and in the half-plane: $\C^+=\left\{z,\, \Im z >0\right\}$ for the determination ${\cal U}^{+}$. Each of the determinations is therefore naturally defined in the respective domains $\C^{\pm}$. Consequently, the functions ${\cal U}^+$ and ${\cal U}^-$ are not analytic on the entire complex plane and are defined on a two-sheeted Riemann surface. The point 0 is a branch point and the positive real axis is a cut-line for these functions.

In view of understanding the behavior of the system when the pole moves from one sheet to another, the problem is then to obtain analytic extensions of ${\cal U}^{\pm}$ to the entire complex plane $\C$. In particular, we want to understand what happens when the pole crosses the real axis.

%Keeping in mind the well-known relation (where the equality is to be understood in the Schwartz distributions meaning): $\frac{1}{x\pm i 0}=PV\left(\frac{1}{x} \right)\mp i\pi \delta_x$, 
The expression in the right-hand side of  (\ref{calU1}) (resp. in (\ref{calU2})) makes sense for $\Im z <0$ (resp. $\Im z>0$). Let us denote $\widetilde{\cal U}^{\pm}$ the corresponding functions.
Using Sokhotski-Plemelj equality \cite{plemelj}, it is seen that, for a real frequency $\omega_0$, it holds :
\begin{eqnarray*}
{\cal U}^{+}(\omega_0+i0)=\fint_{\R^+}  \frac{r^{+}(\r,\omega)}{\omega-\omega_0} e^{i \omega t} d \omega + i\pi r^+(\r,\omega_0)  e^{ i \omega_0 t}, \\
{\cal U}^{-}(\omega_0-i0)=\fint_{\R^+}  \frac{r^{-}(\r,\omega)}{\omega-\omega_0} e^{-i \omega t} d \omega - i\pi r^-(\r,\omega_0) e^{- i \omega_0 t}.
\end{eqnarray*}
where $\fint$ denotes the principal value of the integral and $\omega_0 \in \R^+$. These relations can be rewritten:
\begin{eqnarray}\label{axerel}
{\cal U}^{+}(\omega_0+i0)=\widetilde{\cal U}^{+}(\omega_0-i0)+2i\pi r^+(\r,\omega_0) e^{ i \omega_0 t},\\
{\cal U}^{-}(\omega_0-i0)=\widetilde{\cal U}^{-}(\omega_0+i0)-2i\pi r^-(\r,\omega_0) e^{- i \omega_0 t}.
\end{eqnarray}
This suggests to extend the function analytically in the following way.
We define: 
\begin{eqnarray}
\hbox{For } z \in \C^-, {\cal U}^{+}(z)=\widetilde{\cal U}^{+}(z)+2i\pi r^+(\r,z) e^{ i z t},\\
\hbox{For } z \in \C^+, {\cal U}^{-}(z)=\widetilde{\cal U}^{-}(z)-2i\pi r^-(\r,z) e^{- i z t}
\end{eqnarray}
which extend by analytic continuation ${\cal U}^{+}$ and ${\cal U}^{-}$ to the entire complex plane as entire functions. The behavior on the real axis is regular as can be seen by considering eqs. \ref{axerel}. This means that, even in case 2 (cf. fig.\ref{ComplexPlane} ), the field is well-defined, provided that it is not purely monochromatic, which is of course always the case in any physical situation. In case 3, i.e. when the gain passes in the upper sheet, the field is now exponentially growing in time, which shows that the system cannot be in a stationary state. This is the situation that happens in the quantum theory of the laser when obtaining the number of photons as a function of time \cite{milburn}. Of course, this rapidly leads to a saturation effect that is not taken into account in the present approach. One should implement a formalism such as the master equation \cite[p. 174]{cohen} or the Mawxell-Bloch system \cite[p.222]{brigitte}, to obtain a description not limited to the early times of the phenomenon.

Now the situation at stake with the quantum metamaterial is in fact rather more complicated. Indeed, the gain is not introduced uniformly inside the cavity but is due to the existence of a localized quantum emitter. 
%We aim at describing, at least semi-classically, the interaction of such an emitter with the electromagnetic field. 
The quantum emitter is assumed to have only one radiative transition and to be pumped by an external, classical, field.  The system can be for instance the classical three levels system where the pumping is between $\ket{1}$ and $\ket{3}$, with a non-radiative decay between $\ket{3}$ and $\ket{2}$ and the transition of interest for the emission being $\ket{2}-\ket{1}$.
As a matter of fact, we are interested in the field outside the scatterers, not in the cavity quasi-modes. Those quasi-modes can play the go-between for the continuum of scattering modes and the quantum emitter for resonant scattering. They can also play no role at all if the incoming radiation is not resonant with any quasi-mode. Both situations will be of interest for the applications described in the section devoted to numerical results.
The Hamiltonian of the system is amenable to a Jaynes-Cummings-like hamiltonian, where the pure modes are replaced by the quasi-modes \cite{rotter,rotdicke}.
We develop here an approach based on the scattering matrix and a description of the emitter by mean of the average of its dipolar operator. However, we deal with essentially non-linear phenomena that require the solving of the full system of equations coupling field to matter. %The results given here therefore described the behavior of the system at times smaller than ???????
The existence of the quantum emitter modifies the scattering matrix of the nano-rods and results into the onset of poles at complex frequencies close to the complex resonant frequency of the emitter.

 The first case, that is when the quantum emitter is at resonance with 
the pure cavity mode, corresponds to a situation of strong coupling between the radiative transition of the quantum emitter and the cavity quasi-mode, this leads to damped Rabi oscillations and the formation of polaritons. This will be addressed succinctly in paragraph (\ref{chainpolar}) where a metasurface made of a chain of parallel nanorods is considered. 
%En fait ça, ça ne marche jamais parce que les modes internes sont discrets et complexes donc il faudrait que la fréquence complexe de l'émetteur coïncide exactement avec celle du mode. En revanche pour le mode collectif on peut avoir ce phénomène mais on a alors un polariton délocalisé. 

The second situation where the pure mode is not resonant with the transition of the quantum emitter comes under the weak coupling regime (Purcell effect). The quantum emitter then interacts with a mode belonging to the continuum of scattering modes. When the emitter is driven by an exterior field, it feeds photons inside the nano-rods by stimulated emission.
Describing the quantum emitter by a semi-classical Lorentz model \cite{epsQD}, the scattering matrix of the nano-rod with the emitter can be obtained (see below). However, this calculation can only account for the first moments of the interaction, because it cannot deal with the non-linearity at stake here and therefore the saturation effects are not taken into account. Still, it provides a very convenient way of determining the phase transition towards the emission regime, by considering the path of the pole characterizing the resonance, as explained above. 

\section{Behavior for a collection of scatterers}
\subsection{Derivation of the scattering matrices}
The nano-wires are characterized by a scattering matrix $S(\omega)$ whereas, as shown above, the QOs are small volumes with an effective permittivity $\varepsilon$ with a scattering matrix $S_d(\omega)$. 
Let us consider first the system consisting of a finite set of $N$  nanorods, from the point of view of electromagnetic waves only. This collection of dielectric nanoresonators can support $z$-invariant electromagnetic modes (the $z$ axis being that of the nanorods) with the electric field linearly polarized along $z$. These modes were studied in \cite{strong, imped}. 
%Specifically, there is a continuum of scattering modes and a collection of Bloch modes below the light cone. 
Due to the rotationnal symmetry of the nanorods, the angular momemtum is a good quantum number. 
More precisely, let us denote $\ket{p}= H_p^{(1)}(k_0 r) e^{i p \theta}$ with  $k_0=\omega/c$ and $p \in \N$. Then, under the illumination of an incident monochromatic field $u^i$, the scattered field reads in local coordinates:
\bq
u^s=\sum^N_{m =1} \sum_p s_p^m \ket{p}.
\eq
where $m$ denumbers the nanorods.

%If the incident field is chosen to be a plane wave in the form $u^i=e^{i \alpha x-\beta y}$, where : $\alpha=k_0 \sin \varphi$ and $\alpha^2+\beta^2=k_0^2$, then it holds, by translational invariance of the system: $s^p_n=e^{i\alpha pd} s^0_n $, from which it follows:
%
%$$
%u^s=\sum_n  s^0_n \sum_p e^{i\alpha pd} \ket{p}
%$$
The scattering coefficients $s_p^m$ characterize entirely the system.
Expanding the incident field in the form: $u^i=\sum_l i_l \ket{l}^{\rm reg}$, where $\ket{l}^{\rm reg}=J_l(k_0 r) e^{il\theta}$ and $l \in \N$, it holds, from scattering theory \cite{josaa}: 
\bq
(s^n)=S^n \left[ I^n(i)+ \Sigma_m T^n_m(s^m)\right]
\eq
where $(i)$ denotes the collection of coefficients $i_n$, $(s^n)$ denotes the collection of coefficients $s^n_p,\, p \in \Z$, $T^n_m$ represents the translation from rods $m$ to rod $n$ for the scattering coefficients and $I^n$ the translation for rod $n$ of the incident field coefficients.
%where $\Sigma=\sum_{p\neq 0}  e^{i\alpha pd} T_{0p}$. 
%Therefore, the coefficients $s^n$
%are obtained by the the following relation:
%\bq
%s^=\left[1-S\Sigma \right]^{-1} S 
%\eq
The modes of the system are, by definition, solutions to Maxwell equations in the absence of an incident field.
They are therefore obtained by solving the above system when setting $(i)=0$.
It was shown in \cite{strong} that these modes exist in the vicinity of the Mie resonances and lead to very flat bands, below the light cone.

In order to control the opening of a conduction band, we now insert quantum oscillators (QOs) inside the nanorods. The QOs are described semi-classically as small obstacles having a permittivity of the form \cite{epsQD}
\begin{equation}\label{epsqd}
\varepsilon_{QO}=\varepsilon_b\pm \frac{f}{\omega^2-\omega_0^2+i\omega \Gamma_0}
\end{equation}
The sign $\pm$ indicates whether the QOs are in the absorption or the emission (inversion) regime.

\subsection{Control of the opening of a conduction band}\label{numerical}

\begin{figure}
  \includegraphics[width=10cm]{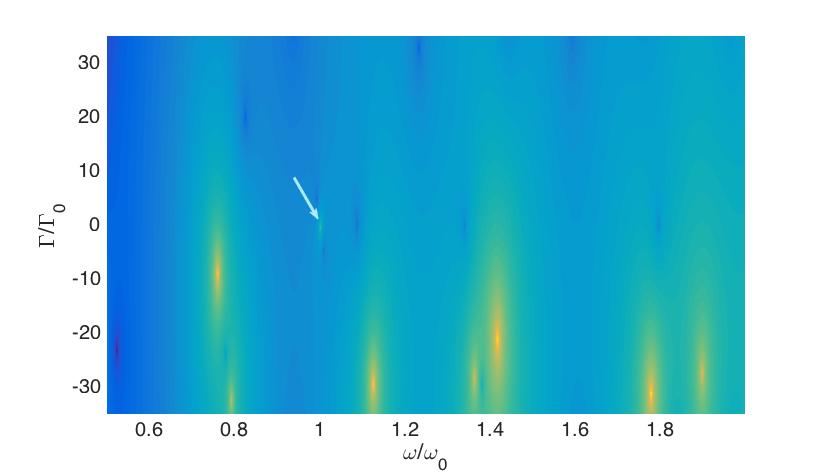}
  \caption{\label{1rod} 
    Modulus of the determinant of the scattering matrix of one rod as a function of the complex frequency $\Omega=\omega+i\Gamma$. The intense points correspond to the poles. The arrow indicates the poles due to the presence of a quantum oscillator. The other poles correspond to Mie resonances.}
\end{figure}

\begin{figure}
  \includegraphics[width=10cm]{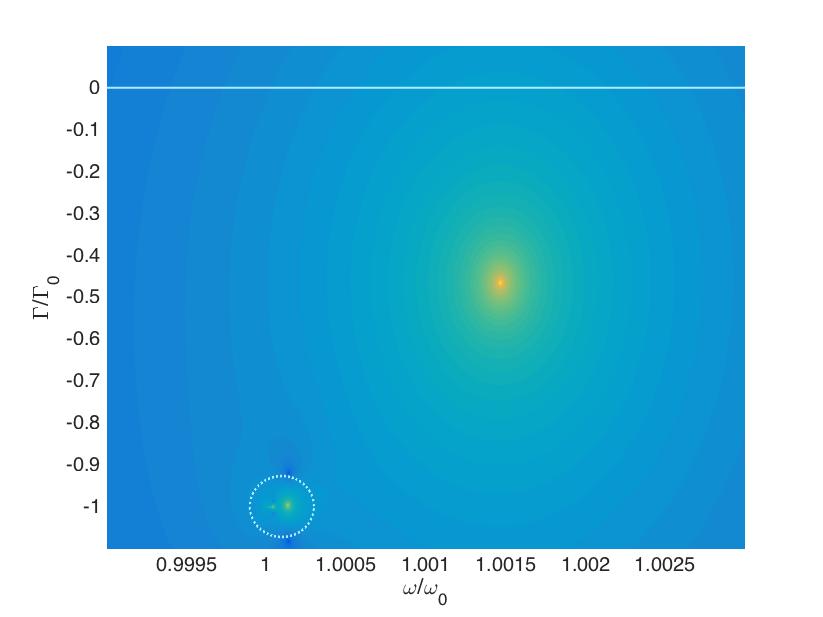}
  \caption{\label{1rodzoom} 
    Zoom on the fig. \ref{1rod} around the poles due to the QO. The light blue line indicates the real axis. The light blue circle encircles the pole structure due to the pole in the effective permittivity (cf. eq. (\ref{epsqd})). The light blue line indicates the real axis.}
\end{figure}
\begin{figure}
  \includegraphics[width=10cm]{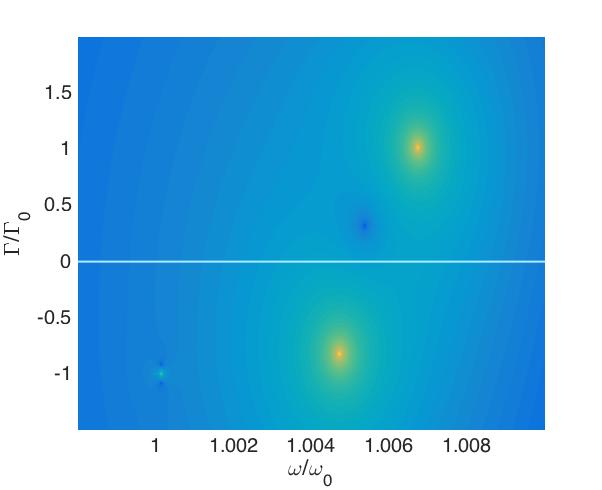}
  \caption{\label{2rods} 
    Modulus of the determinant of the scattering matrix for two rods with a QO inserted, close to each other. The intense points correspond to the poles. One of the pole is the upper part of the complex plane, indicating a super-radiant state. The second pole in the lower part corresponds to a sub-radiant state. The light blue line indicates the real axis.}
\end{figure}
\begin{figure}
  \includegraphics[width=10cm]{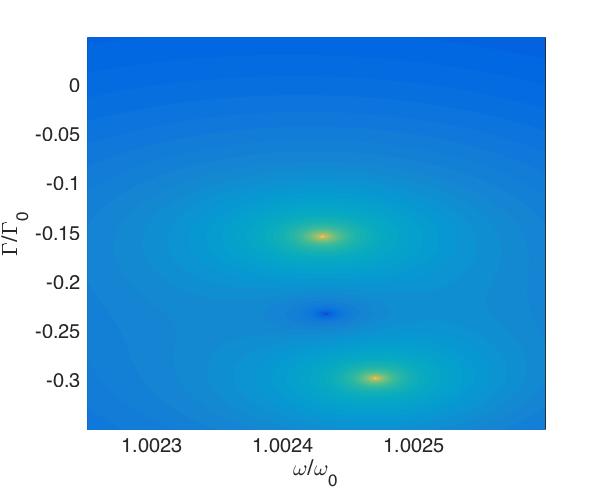}
  \caption{\label{2rodsub} 
    Modulus of the determinant of the scattering matrix for two rods far apart. The intense points correspond to the poles. There are two poles, both being in the lower part, indicating that the light emitting transition was not reached.}
\end{figure}

We start with a single scatterer in which a quantum emitter is embedded, a structure that we call a hybrid scatterer. The nanorods are circular with a relative permittivity of 12. The transition of the QOs is set to $\lambda_0=450 {\rm nm}$ (i.e. with a corresponding pulsation of $\omega_0=4.188\cdot 10^{15} {\rm rad/s}$), the inverse lifetime is $\Gamma_0=7.85 \cdot 10^{12} {\rm rad/s}$. The pole structure is shown in fig. \ref{1rod} and fig. \ref{1rodzoom}. There the modulus of the determinant of the scattering matrix is represented, as a function of the complex frequency $\Omega=\omega+i \Gamma$. The horizontal axis is normalized to $\omega_0$ and the vertical axis to $\Gamma_0$. The modulus of the determinant of the scattering matrix is represented. The horizontal abscissa corresponds to the real part of the frequency, normalized to the transition of the QO, while the vertical abscissa is the imaginary part of the frequency, that is the inverse of the life-time of the transmission, normalized to that of the QO. There it can be seen that there is a pole of the hybrid scatterer (fig. \ref{1rodzoom} is a zoom on the region  of interest), resulting from the coupling between the emitter and the continuum of scattering states as well as a concentration of poles due to the pole of the effective permittivity of the QO (circled with a dashed blue line in fig. \ref{1rodzoom}), given in eq.\ref{epsqd}. The latter are not of interest to us. 
Although there is gain in the system, the state of the hybrid system has a finite lifetime, because the radiative losses overcome the gain, therefore the pole of the hybrid state is in the lower part of the complex plane. The situation is quite different when two hybrid scatterers are close enough to each other to interact: in fig.\ref{2rods} the pole structure of a set of two hybrid scatterers is given. One can see the onset of two poles. When the scatterers are close enough, one of the pole crosses the real axis and enters the second sheet: the system is then in a regime of light amplification, with the same {\it caveats} already made above. This shows the onset of a collective behavior similar to the super-radiance and sub-radiance effects that exist in the simpler context of two atoms interacting via a continuum \cite[p. 585]{cohen}. In this situation, two states appear, one of which as a shorter radiative life-time than that of an isolated atom, it is then called a super-radiant state. The other one, on the contrary is stable and is called sub-radiant. In our situation, the super-radiant state is related to the pole in the upper part of the complex plane, while the sub-radiant states are situated below.  This super-radiant state will prove to be crucial for the control of conduction bands. Note that, when the rods are too far apart, the quasi-modes remain in a subradiant state (fig. \ref{2rodsub})

\begin{figure}
  \includegraphics[width=10cm]{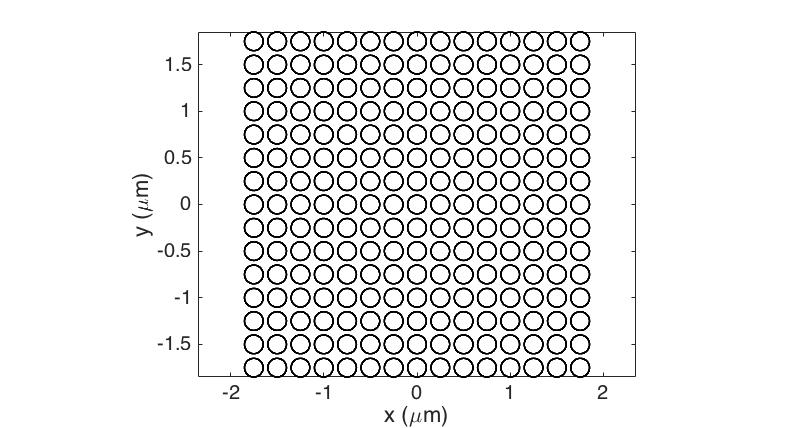}
  \caption{\label{meta} A few cells of the quantum metamaterials. The nanorods are disposed along a square lattice.}
 \end{figure}

\begin{figure}
  \includegraphics[width=10cm]{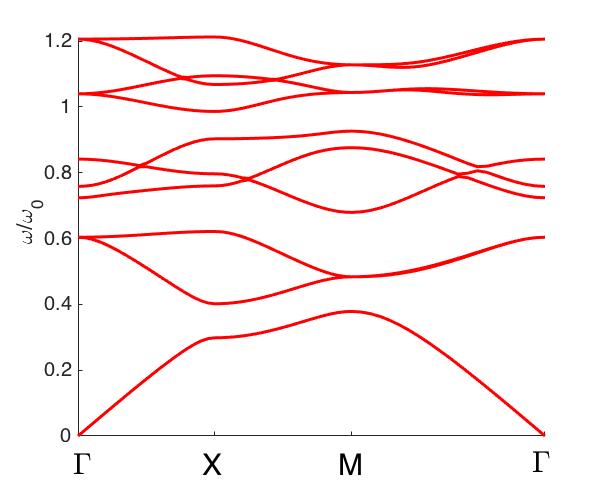}
  \caption{\label{bloch} 
    Bloch diagram for an infinite bidimensional structure of bare nanorods. A few cells are represented in fig. \ref{meta}.}
\end{figure}

We now increase the number of scatterers while positioning them in such a way that the bare structure, that is without QOs embedded, presents a photonic band gap. This is obtained by considering a bidimensional periodic set of scatterers such as that depicted in fig.\ref{meta}. 
In the case of an infinite periodic structure, the waves present a Bloch spectrum given in fig. \ref{bloch}.
   
\begin{figure}
  \includegraphics[width=10cm]{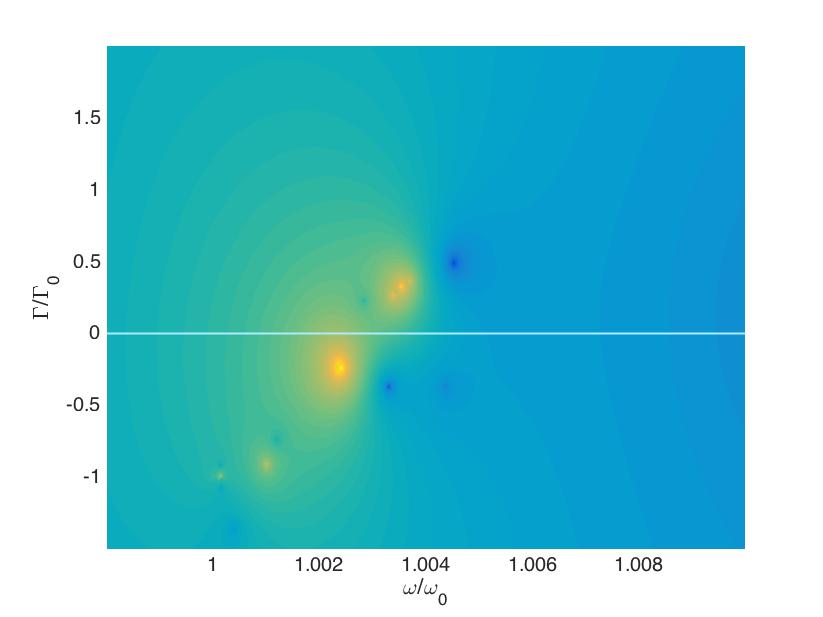}
  \caption{\label{3rod} 
    Modulus of the determinant of the scattering matrix as a function of the complex frequency $\Omega=\omega+i\Gamma$ for the 3x3 nanorods, with the QOs in the emission regime. There is a concentration of poles in the upper complex plane due the coupling between the rods.}
\end{figure}

The pole structure of the 3x3 nanorods structure is given in fig.\ref{3rod}. Because of the coupling between the rods, there is now a cluster of poles above the real axis. As the number of nanorods increases, this accumulation provokes the onset of a conduction band, as will be shown in the following.

\begin{figure}
  \includegraphics[width=10cm]{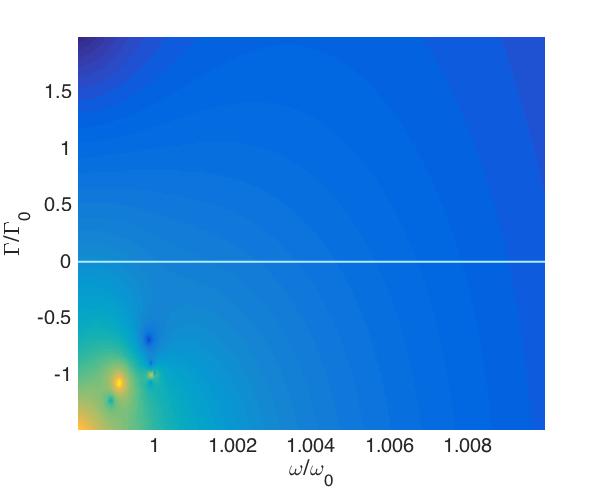}
  \caption{\label{3rodabs} 
    Modulus of the determinant of the scattering matrix as a function of the complex frequency $\Omega=\omega+i\Gamma$ for a set of 3x3 rods with the QOs in the absorption regime.}
\end{figure}

For completeness and the sake of comparison, the pole structures of the 3x3 structure in the absorption regime, i.e. without inversion, is given in fig. \ref{3rodabs}.

\begin{figure}
  \includegraphics[width=10cm]{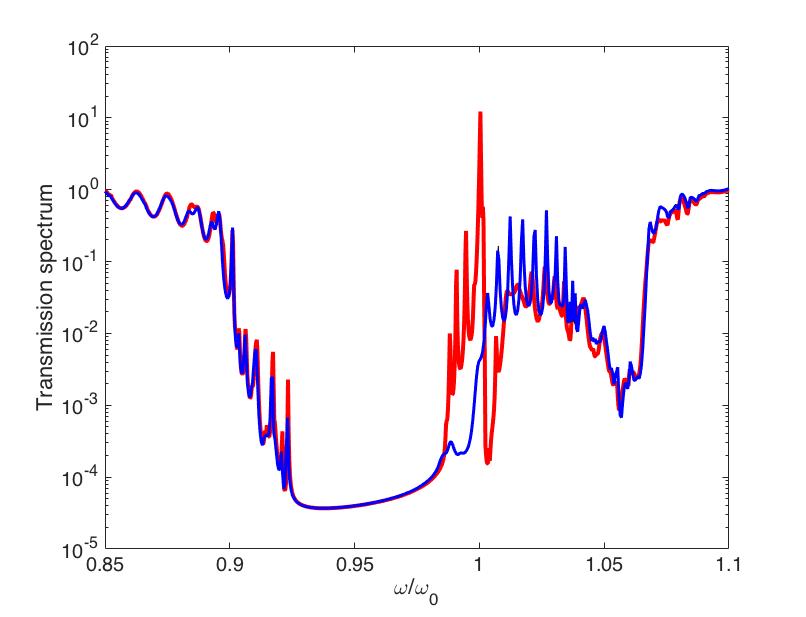}
  \caption{\label{trans} 
   Transmission spectrum for a $15\times 15$ rods. The curve in blue corresponds to the QOs in the absorption regime, the red curve to the QOs in the emission regime.}
\end{figure}

We are interested in the possibility of inducing a conduction band in the framework  of a pump-probe experiment. We choose the radiative transition of the QO to lie in the forbidden band situated in the normalized interval $\sim [0.8, 1]$ (this corresponds to a wavelength in vacuum of $450$ nm). It is to be remarked that this region of frequencies does not correspond to a full band gap, rather it is a region where the Bloch modes cannot be excited for a normal incidence. As we use a finite structure, this leads to a low, but not null, density of states. The transmission spectrum of the structure with quantum dots embedded, but without a pump field, is given in fig. \ref{trans} (red curve). When the pump field is switched on (blue curve), the quantum dots are in the emission regime and the transmission spectrum corresponds to the continuous curve. There peaks can be seen, corresponding to a conduction band created by the pumped QOs.
This shows the possibility of controlling the opening and closing of a region of high transmission by purely optical means.
\subsection{Remarks on the behavior of an infinite chain of scatterers}\label{chainpolar}

We consider here an infinite periodic chain of hybrid resonators. This situation was already alluded above. The system of nanorods and QOs is described by the following hamiltonian (because of the periodicity of the amplitude of the mode, note $V$ depends only upon the position $r_0$ of the QO inside the basic cell):
\bq
\begin{array}{r}
H=\int_0^{+\infty} dk \hbar \omega(k) a(k)^+a(k)
+\sum_{n} \hbar \omega_n a^+_{n} a_{n} \\
+\frac{1}{2} \Omega \sum_n \sigma^n_z\\
+\sum_{n,p} V_{p} (a_p^++a_p)(\sigma^n_++\sigma^n_-)
\end{array}
\eq
The Hamiltonian is acting on the Fock space: ${\cal F}=\C^n \otimes {\cal F}_{\rm bosonic}$.

This hamiltonian can be further simplified by noting that a quantum emitter resonant with the modes, as the ones we consider here, will preferentially couple to the modes guided along the chain. 

%This is illustrated in fig. \ref{fig2}, where the situated of a classical dipole in the vicinity of the chain is investigated. Insérer le numérique ici

The electromagnetic Bloch modes is characterized by the Boch wavevector $k\in Y^*$, being a good quantum number.
The Hamiltonian reads as:
\bq
H=\sum_{n} \hbar \omega_n a^+_{n} a_{n}+\frac{1}{2} \Omega \sum_n \sigma^n_z+ \sum_p (a_p^++a_p) V_p \sum_n e^{ikn}(\sigma^n_++\sigma^n_-)
\eq
then we can define the Wannier transform of the moment operators: $J_{\pm}(k) = \sum_n e^{ikn} \sigma^{n}_{\pm}$.
We are particularly interested in the modes linked to the Mie resonances, which turn out to be quite flat, being essentially defect modes (the underlying collective behavior was analyzed in \cite{PRLVynck} and the modes in \cite{strong}). The physics at stake here can therefore be analyzed by setting $k=0$, i.e. by considering periodic fields. Assuming the above and that the QO is resonant with one of the modes %and the system is illuminated by a classical external field $u_e$
, the hamiltonian is therefore simplified to the following Dicke form $H_{\rm eff}$ :
%$$
%H_{\rm int}=V(a^++a) \left(  \sum_n \sigma^{2n}_{\pm} -  \sum_n \sigma^{2n+1}_{\pm} \right)
%$$
%this suggests to define the collective (partial) creation operators:
%$$
%J^o_{\pm}=\sum_n \sigma^{2n+1}_{\pm},\,  J^e_{\pm}=\sum_n \sigma^{2n}_{\pm}
%$$
%and
%$$
%J^o_{x}=\sum_{\pm} \, J^o_{\pm},\, J^e_{x}=\sum_{\pm} \, J^e_{x}
%$$
%as well as $J^{e,o}_z=\sum_n \sigma^{2n,2n+1}_z$, $J_{\pm}=J^e_{\pm}+J^o_{\pm}$ and $J_x=J^e_x+J^o_-$
\bq
H_{\rm eff}=\hbar \omega a^+ a+\frac{1}{2} \Omega J_z+V (a^++a)\, (J_++J_-),%+V_0 \, \cos(\omega t) \, (J_++J_-)
\eq

%\bq
%H_{\rm eff}=\hbar \omega a^+ a+\frac{1}{2} \Omega J_z+V (a^++a)\, \left(J^e_x-J^o_x\right)+V_0 \, \cos(\omega t) \, \left(J^e_x-J^o_x\right)
%\eq
%This leads to two sets of quantum states: $\ket{j^o,m^o}$ and $\ket{j^e,m^e}$ assuming the maximum value for $j^o=N_o/2$ and $j^e=N_e/2$
where the operators $J_z,\, J_{\pm}$ satisfy the $su(2)$ Lie algebra $[J_z,J_{\pm}]=\pm J_{\pm}$ and $[J_+,J_-]=2J_z$.

%Let us develop the scattering theory corresponding to $(H_{\rm eff},H_0)$ where $H_0=\hbar \omega a^+ a+\frac{1}{2} \Omega J_z$ is the free hamiltonian. 
Realizing the coupling between the Bloch modes that can exist along the chain and the excitations of the QOs would result in the hybrid states named polaritons \cite{polariton}. Theses quasi-particles are different from the usual cavity polaritons \cite{kavokin} in that they are delocalized in space along the chain: they result from the coupling between the QOs through the Bloch mode.
%Let us prepare an incoming state Fock state $ \ket{-} \otimes \ket{0}$ corresponding to the QOs in the excited states and no photon in the quantum field.
This suggests that it should be possible to observe in that situation a superradiant quantum phase transition \cite{dickeQPT}. The semi-classical analysis developed above shows clear indication towards that direction.

%Let us develop thescattering theory corresponding to $(H_{\rm eff},H_0)$ where $H_0=\hbar \omega a^+ a \int_{\omega} \rho(\omega) a^+_{\omega}a_{\omega} d\omega+\frac{1}{2} \Omega \sigma_z$ is the free hamiltonian and the interaction term is dipolar. We use an approach where 
\section{conclusion}
We have proposed a quantum metamaterial, made of a collection of dielectric nanorods in which a quantum oscillator is embedded. Working in the frame work of the linear response theory and the non-perturbative resolvent approach, the existence of a pole structure of the scattering matrix was identified. This allows to predict the lasing transition in the system and the possibility of a tunable conduction band, in the framework of a pump-probe experiment. The results also suggest the existence of a quantum phase transition in the regime of strong coupling. This could be addressed by developing a full quantum approach to the interaction of the electromagnetic field with the quantum oscillators.


\begin{thebibliography}{0}

  \bibitem{capolino}% 
  \textsc{F. Capolino ed.}, 
 Theory and Phenomena of Metamaterials (CRC Press, London, 2009).

  \bibitem{review}% 
  \textsc{N. I. Zheludev,}
   \textsc{Y. S. Kivshar},
  Nat. Mat. \textbf{11}, 917 (2012).
  
  \bibitem{metasurf}
  \textsc{C. L. Holloway,}
  \textsc{E. F. Kuester,} 
  \textsc{J. A. Gordon,}
  \textsc{J. O'Hara,} 
  \textsc{J. Booth,}
   \textsc{D. R. Smith},
   IEEE Antenn. and Propag. \textbf{54}, 10 (2012).
  
  
  \bibitem{gain}% 
  \textsc{O.~Hess,}
  \textsc{K. L.~Tsakmakidis}, 
  Science \textbf{339}, 654 (2013).
  
   \bibitem{metaQD}% 
  \textsc{M.~Decker et al.},
  Nat. Comm. \textbf{4}, 2949 (2013).
  
   \bibitem{rewalex}% 
  \textsc{A.~Zagoskin,}
    \textsc{E.~Rousseau,}
  \textsc{D.~Felbacq},
  EPJ : Quantum Technologies \textbf{3}, 2 (2016). 
  
  \bibitem{moiQ}% 
  \textsc{D.~Felbacq},
  J. Nanophoton. \textbf{5}, 050302 (2011).
  
   
   \bibitem{Qmat} 
     \textsc{A. L. Rakhmanov,}
  \textsc{A.~Zagoskin,}
  \textsc{Sergey Savel?ev,}
  \textsc{F.~Nori},
  Phys. Rev. B \textbf{77}, 144507 (2008).
  
   \bibitem{Qmat2} 
     \textsc{A. Shvetsov,}
  \textsc{A. M. Satanin,}
  \textsc{S. Savel'ev,}
  \textsc{F.~Nori,}
    \textsc{A.~Zagoskin},
  Phys. Rev. B \textbf{77}, 144507 (2008).
  
  \bibitem{manipQmat}% 
  \textsc{P.~Macha et al.},
  Nat. Comm. \textbf{5}, 5146 (2014).
  
  \bibitem{reedsimon}% 
  \textsc{M. Reed,} 
  \textsc{B. Simon},
  Methods of Modern Mathematical Physics, I: Functional Analysis (Academic Press, New York, 1980).
  \bibitem{milburn}% 
  \textsc{D. F. Walls,} 
  \textsc{F. G. Milburn},
   Quantum Optics (Springer-Verlag, Berlin, 1994).

 \bibitem{carmichael}% 
  \textsc{H. Carmichael} ,
An Open Systems Approach to Quantum Optics (Springer-Verlag, Berlin, 1993).

\bibitem{cohen}% 
  \textsc{C. Cohen-Tannoudji,}
     \textsc{J. Dupont-Roc,}
       \textsc{G. Grynberg},
Atom-Photon Interactions (Wiley-VCH, Weinheim, 2004).

\bibitem{rotter}% 
  \textsc{I. Rotter},
  J. Phys. A: Math. Theor. \textbf{42}, 153001 (2009).

  \bibitem{collision}% 
  \textsc{M. L. Goldberger,} 
  \textsc{K. M. Watson},
  Collision Theory (John Wiley \& Sons, New York, 1964).

  \bibitem{melrose}% 
  \textsc{R. Melrose}, 
 Geometric Scattering Theory (Cambridge University Press, Cambridge, 1995).

 \bibitem{moipole}% 
  \textsc{D. Felbacq,}    
  \textsc{R. Sma\^ali},
  Phys. Rev. B   \textbf{67}, 085105 (2003).
  
 \bibitem{moipole2}% 
  \textsc{D. Felbacq} ,   
  Phys. Rev. E   \textbf{64}, 047702 (2001).

\bibitem{simon}% 
  \textsc{B. Simon},    
  Quantum Chemistry   \textbf{14}, 529 (1978).

\bibitem{moiseyev}% 
  \textsc{N. Moiseyev} ,   
  Phys. Rep.   \textbf{302}, 211 (1998).

  \bibitem{bohm}% 
  \textsc{A. Bohm,} 
  \textsc{M. Gadella},
  Dirac Kets, Gamow Vectors and Gel'fand Triplets (Springer-Verlag, Berlin, 1969).

\bibitem{hugonin}
\textsc{C. Sauvan,}
\textsc{J. P. Hugonin,}
\textsc{I. S. Maksymov,} 
\textsc{P. Lalanne},
Phys. Rev. Lett. \textbf{110}, 237401 (2013). 


\bibitem{soukoulis}%
 \textsc{X. Jiang,}
  \textsc{Q. Li,}
  \textsc{C. M. Soukoulis},    
  Phys. Rev. B: Rapid Comm.   \textbf{59}, R9007 (1999).
  
  \bibitem{plemelj}% 
  \textsc{J. Plemelj} , 
Problems in the sense of Riemann and Klein (Interscience Publishers, New York, 1964).


  \bibitem{brigitte}% 
  \textsc{B. Bidegaray-Fesquet}, 
 Hiérarchie de modèles en optique quantique (Springer-Verlag, Berlin, 2006).

\bibitem{rotdicke}% 
  \textsc{H. Eleuch,}   
  \textsc{I. Rotter},
  Int.J.Theor.Phys.   \textbf{54}, 3877 (2015).

  \bibitem{epsQD}% 
  \textsc{P. Holmstr\"om,}   
  \textsc{L. Thyl\'en,}
  \textsc{A. Bratkovsky},
  J. Appl. Phys.  \textbf{107}, 064307 (2010).
  
   \bibitem{strong}% 
  \textsc{D. Felbacq} ,  
  Superlatt. and Microstruct.  \textbf{78}, 79 (2015).
  
   \bibitem{imped}% 
  \textsc{D. Felbacq},   
  Math. Probl. in Engineer.   \textbf{2015}, 473079 (2015).
  
  \bibitem{josaa}% 
  \textsc{D. Felbacq,}
  \textsc{G. Tayeb,}
  \textsc{D. Maystre},
  J. Opt. Soc. Am. A   \textbf{11}, 2526 (1994).
  
  \bibitem{PRLVynck}
  \textsc{K. Vynck et al.},
  Phys. Rev. Lett.  \textbf{102}, 133901  (2009).
  
  
  \bibitem{polariton}
  \textsc{D. Sanvitto,}
  \textsc{S. Kéna-Cohen},
  Nat. Mat. \textbf{15}, 1061 (2016).
  
  
  \bibitem{kavokin}% 
  \textsc{A. Kavokin,} 
  \textsc{G. Malpuech}, 
 Cavity Polaritons (Elsevier, Amsterdam, 2003).
 
   \bibitem{dickeQPT}% 
  \textsc{C. Emary,}
  \textsc{T. Brandes} ,  
  Phys. Rev. E   \textbf{67}, 066203 (2003).
   
\end{thebibliography}
\end{document}